\documentclass[12pt]{article}

\newcommand{\be}{\begin{equation}}
\newcommand{\ee}{\end{equation}}
\newcommand{\bi}[1]{\vspace{-3mm} \bibitem{#1}}
\usepackage{epsfig,amsmath,amssymb,graphics,graphicx}

\begin{document}

\begin{center}
Physics Letters A 299 (2002) 173-178.

\vskip 5 mm
{\Large \bf Stationary States of  \\
Dissipative Quantum Systems}
\vskip 5 mm
{\Large \bf Vasily E. Tarasov } \\

\vskip 3mm
{\it Skobeltsyn Institute of Nuclear Physics, \\
Moscow State University, Moscow 119991, Russia}

{E-mail: tarasov@theory.sinp.msu.ru}
\end{center}

\vskip 5 mm
\begin{abstract}
In this Letter we consider stationary states of
dissipative quantum systems. We discuss stationary states of
dissipative quantum systems, which coincide with stationary states
of Hamiltonian quantum systems. Dissipative quantum systems with pure
stationary states of linear harmonic oscillator are suggested.
We discuss bifurcations of stationary states for dissipative quantum
systems which are quantum analogs of classical dynamical bifurcations.

\end{abstract}

\vskip 5mm
PACS {03.65.-w; 03.65.Yz}

\vskip 5mm
Keywords: Dissipative quantum systems, Stationary states, Bifurcation


\section{Introduction}


The dissipative quantum systems are of
strong theoretical interest \cite{W1}. 
As a rule, any microscopic system is always embedded in some 
(macroscopic) environment and therefore it is never really isolated. 
Frequently, the relevant environment is in principle
unobservable or it is unknown \cite{Mash,T1}.
This would render theory of dissipative quantum systems
a fundamental generalization of quantum mechanics \cite{Prig}.

Spohn \cite{Spohn,Spohn2,Spohn3} derives sufficient condition for
existence of an unique stationary state for dissipative quantum
system described by Lindblad equation. 
The irreducibility condition given by \cite{Dav1} defines 
stationary state of dissipative quantum systems. 
An example, where the stationary state is unique and
approached by all states for long times is considered by Lindblad
\cite{Lind2} for Brownian motion of quantum harmonic oscillator.
The stationary solution of Wigner function evolution equation
for dissipative quantum system was discussed in \cite{AH,ISS}.
Quantum effects in the steady states of  the
dissipative map are considered in \cite{DC}.

\section{Definition of stationary states}

In the general case, the time evolution of quantum state $\rho_t$ is
described by Liouville-von Neumann equation 
\be \label{LN}
\frac{d}{dt}  \rho_t= \hat \Lambda  \rho_t ,  \ee 
where $\hat \Lambda$ is a quantum Liouville operator. 
For Hamiltonian systems quantum Liouville operator has the form 
\be \label{hLo}  
\hat \Lambda \rho_t= -\frac{i}{\hbar}[ H, \rho_t] , \ee 
where $H=H(q,p)$ is a Hamilton operator. 
If quantum Liouville operator
$\hat \Lambda$ cannot be represented in the form (\ref{hLo}), then
quantum system is called non-Hamiltonian or dissipative quantum system.
Stationary state is defined by the condition
\[ \hat \Lambda \rho_t=0 . \]
For Hamiltonian systems this condition has the form
\be \label{s21} [ H , \rho_t ]=0 . \ee

\section{Pure stationary states of Hamiltonian systems}

A pure state $\rho_{\Psi}=|\Psi><\Psi|$ is a stationary state of
Hamiltonian quantum system, if $|\Psi>$ is an eigenvector of
Hamilton operator $H=H(q,p)$.
Using $<\Psi|\Psi>=1$, we get the equality (\ref{s21}) in the form
\be \label{HE} H|\Psi>=|\Psi>E , \ee
where $E=<\Psi| H|\Psi>$.
Equation (\ref{HE}) defines pure
stationary states $|\Psi>$ of Hamiltonian systems.
Eigenvalues of Hamilton operator are identified with the energy of
the system. It is known, that Hamilton operator for linear
harmonic oscillator is
\be \label{harm}
H= \frac{ p^{2}}{2m}+\frac{m \omega^{2} q^{2}}{2} . \ee
Equation (\ref{HE}) has the solution if \be \label{En}
E_{n}=\frac{1}{2}\hbar\omega(2n+1) . \ee 
In coordinate
representation stationary states of linear harmonic oscillator are
\be \label{Psi} \Psi_n(q)= \frac{1}{q_0} exp
\Bigl(-\frac{q^2}{2q^2_0}\Bigr) {\cal H}_n \Bigl(\frac{q}{q_0}
\Bigr) , \quad q_0=\sqrt{\frac{\hbar}{m \omega}} , \ee 
where ${\cal H}_n({q}/{q_0})$ is Hermitian polynomial of order $n$.

\section{Pure stationary states of dissipative systems}

Let us consider Liouville-von Neumann equation (\ref{LN}) of the form
\be \label{h1} 
\frac{d}{dt} \rho_{t}=-\frac{i}{\hbar}[ H, \rho_{t}] +
\sum^s_{k=1}  \hat F_k N_k(\hat L_H,\hat R_{H})  \rho_{t} . \ee
Here $\hat F^{k}$ are operators act on operator space, $\hat L_A$ and
$\hat R_A$ are operators of left and right multiplication 
\cite{Tarmsu} defined by
\[ \hat L_A B=AB , \quad \hat R_A B=BA , \]
for all operators $B$.

Let $\rho_{\Psi}=|\Psi><\Psi|$ is a pure state with eigenvector
$|\Psi>$ of the Hamilton operator $H$.
If equation (\ref{HE}) is satisfied, then the state
$\rho_{\Psi}=|\Psi><\Psi|$
is a stationary state of Hamilton system
\be \label{h2} \frac{d}{dt} \rho_{t}=-
\frac{i}{\hbar}[ H, \rho_{t}] , \ee
associated with dissipative system (\ref{h1}).

If the vector $|\Psi>$ is eigenvector of $H$, then Liouville-von
Neumann equation (\ref{h1}) for pure state
$\rho_{\Psi}=|\Psi><\Psi|$ has the form
\[ \frac{d}{dt} \rho_{\Psi}=
\sum^s_{k=1} N_{k}(E,E) \hat F_{k} \rho_{\Psi} , \]
where the functions $N_k(E,E)$ are defined by
\[ N_{k}(E,E)=  
<\Psi|(N^{\dagger}_{k}(\hat L_H,\hat R_{H})  I)|\Psi> . \]
Operator $N^{\dagger}_{k}(\hat L_{H},\hat R_{H})$ is 
adjoint operator on operator space defined by
\[ (N^{\dagger}_{k}(\hat L_{H},\hat R_{H})A|B)=
(A|N_{k}(\hat L_{H},\hat R_{H})B) , \]
where $(A|B)=Tr(A^{\dagger}B)$.
If all functions $N_{k}(E,E)$ are equal to zero
\be \label{sc} N_{k}(E,E)=0 , \ee
then the stationary state of Hamiltonian quantum system (\ref{h2})
is stationary state of dissipative quantum system (\ref{h1}).

Note, that functions $N_{k}(E,E)$ are eigenvalues and $|\Psi>$ is
eigenvector of operators 
$N_{k}(H,H)= N^{\dagger}_{k}(\hat L_{H},\hat R_{H}) I$, 
since
\[ N_{k}( H, H)|\Psi>=|\Psi>N_{k}(E,E) . \]
Therefore stationary states of dissipative quantum system
(\ref{h1}) are defined by zero eigenvalues of operators 
$N_k(H,H)=N^{\dagger}_{k}(\hat L_{H},\hat R_{H}) I$.

\section{Dissipative systems with oscillator stationary states}

In this section we consider simple examples of dissipative
quantum systems (\ref{h1}).

1) Let us consider nonlinear oscillator with friction defined by
the equation 
\be \label{nlo1} 
\frac{d}{dt}  \rho_t= -\frac{i}{\hbar}[ H_{nl},  \rho_t] + 
\frac{i}{\hbar}\beta  [ q^2, p^2 \circ  \rho_t] , \ee 
where Hamilton operator $H_{nl}$ is
\[ H_{nl}=\frac{ p^2}{2m}+
\frac{m \Omega^2  q^2}{2}+\frac{\gamma  q^4}{2} , \] 
and
\[ A\circ B=\frac{1}{2}(AB+BA) . \]
Equation (\ref{nlo1}) can be rewritten in the form 
\be \label{nlo2}
\frac{d}{dt} \rho_t= -\frac{i}{\hbar}[ H, \rho_t] 
+ \frac{2im\beta}{\hbar} [ q^2, \ (\frac{ p^2}{2m}+\frac{\gamma q^2}{2m\beta}
- \frac{\Delta}{4\beta} I ) \circ  \rho_t] , \ee
where $\Delta=\Omega^2-\omega^2$, and $H$ is Hamilton operator of
linear harmonic oscillator (\ref{harm}).
Equation (\ref{nlo2}) has the form (\ref{h1}), where
\[ \hat F =\frac{2im\beta}{\hbar} (\hat L_{q^2}-\hat
R_{q^2}) , \]
\[ N(\hat L_H,\hat R_H)=\frac{1}{2}(\hat L_H+\hat
R_H)- \frac{\Delta}{2\beta}\hat L_{I} , \]
\[ N(E,E)=<\Psi| H-\frac{\Delta}{2 \beta} I|\Psi>=
E-\frac{\Delta}{2 \beta} . \] Let $\gamma=\beta m^2\omega^2$. 
The dissipative system (\ref{nlo1}) has one stationary state
(\ref{Psi}) of harmonic oscillator with energy 
$E_n=(\hbar \omega /2)(2n+1)$, if
\[ \Delta=2\beta \hbar \omega (2n+1) , \]
where $n$ is an integer nonnegative number. 
This stationary state is one of  stationary states of linear harmonic
oscillator with the mass $m$ and frequency $\omega$. 
In this case, we can have the quantum analog \cite{Tarpla} 
of dynamical Hopf bifurcation \cite{TL,MMC}.

2) Let us consider dissipative system described by evolution equation
\be \label{h4} 
\frac{d}{dt} \rho_{t}=-\frac{i}{\hbar}[ H, \rho_{t}]
+\frac{i}{\hbar} [q, N(\hat L_H,\hat R_{H}) \rho_{t}] , \ee
where the Hamilton operator is defined by (\ref{harm}) and
\be \label{h4n}
N(\hat L_H,\hat R_{H})=cos\Bigl(
\frac{\pi}{2\varepsilon_0} (\hat L_{H}+\hat R_{H})\Bigr) 
=\sum^{\infty}_{m=0}\frac{1}{(2m)!} \Bigl(
\frac{i\pi}{2\varepsilon_0}\Bigr)^{2m} (\hat L_{H}+\hat
R_{H})^{2m} . \ee
The operator $\hat F$ on operator space is
\[ \hat F=\frac{i}{\hbar}(\hat L_q-\hat R_q) . \]
The function $N(E,E)$ has the form
\[ N(E,E)=cos\Bigl( \frac{\pi E}{\varepsilon_0}\Bigr)=
\sum^{\infty}_{m=0}\frac{1}{(2m)!}
\Bigl( \frac{i\pi E}{\varepsilon_0}\Bigr)^{2m} . \]
The stationary state condition (\ref{sc}) has the solution
\[ E=\frac{\varepsilon_0}{2} (2n+1) , \]
where $n$ is an integer number.
If parameter $\varepsilon_{0}$ is equal to $\hbar \omega$, 
then quantum system (\ref{h4}), (\ref{h4n}) has stationary states
(\ref{Psi}) with the energy (\ref{En}).
As the result stationary states of dissipative
quantum system (\ref{h4}) coincide with stationary 
states (\ref{Psi}) of the linear harmonic oscillator.

If the parameter $\varepsilon_{0}$ is equal to $\hbar
\omega(2l+1)$, then quantum system (\ref{h4}), (\ref{h4n}) has
stationary states (\ref{Psi}) with $n(k,m)=2kl+k+l$ and
\[ E_{n(k,l)}=\frac{\hbar \omega}{2}(2k+1)(2l+1) . \]

3) Let us consider the operators $N_{k}(\hat L_{H},\hat R_{H})$
in the form
\[ N_{k}(\hat L_{H},\hat R_{H})
=\frac{1}{2\hbar} \sum_{n,m} v_{kn} v^{*}_{km}
( 2\hat L^{n}_{H} \hat R^{m}_{H}-
\hat L^{n+m}_{H}-\hat R^{n+m}_{H} ) , \]
and $\hat F_{k}=\hat L_{I}$. In this case,
Liouville-von Neumann equation (\ref{h1}) has the form of
Lindblad equation \cite{kn2,Lind,AL}:
\be \label{Lin}
\frac{d}{dt} \rho_t=-\frac{i}{\hbar}[H,\rho_t]
+\frac{1}{2\hbar} \sum^{m}_{k=1}([ V_{k} \rho_t, V_{k}^{\dagger} ]
+ [ V_{k}, \rho_t V_{k}^{\dagger} ]). 
\ee
with operators
\[ V_{k}=\sum_{n}v_{kn}H^{n} , \ \ V^{\dagger}_{k}=\sum_{m}v^{*}_{km}H^{m}. \]
If $\rho_{\Psi}=|\Psi><\Psi|$ is a pure stationary state, 
then $N_{k}(E,E)=0$ and this state is a stationary state of the
dissipative quantum system (\ref{Lin}).

\section{Dynamical bifurcations and catastrophes}

Let us consider a special case of dissipative quantum
systems (\ref{h1}) such that
the function $N_{k}(E,E)$ be a potential function,
i.e., we have a potential $V(E)$ such that
\[ \frac{\partial V(E)}{\partial E_{k}}=N_{k}(E,E) , \]
where $E_k=<\Psi| H_k|\Psi>$, and
\[ N_k(E,E)=<\Psi|(N^\dagger_k(\hat L_H,\hat R_H) I)|\Psi> , \]
\[  H=\sum^s_{k=1}  H_k  , \quad  H_k|\Psi>=|\Psi>E_k . \]
In this case, the stationary condition (\ref{sc}) for
dissipative system (\ref{h1})is defined by critical points of
the potential $V(E)$. If the system has one variable $E$, 
then the function $N(E,E)$ is always potential function.
In the general case, the functions $N_{k}(E,E)$ are potential, if
\[ \frac{\partial N_k(E,E)}{\partial E_l}= \frac{\partial
N_l(E,E)}{\partial E_k} . \]
Stationary states of dissipative quantum system (\ref{h1})
with potential functions $N_{k}(E,E)$ are depend on critical
points of potential $V(E)$. 
This allows to use theory of
bifurcations and catastrophes for parametric set of functions
$V(E)$. Note that a bifurcation in a space of variables
$E=\{E_k|k=1,...,s \}$ is a bifurcation in the space of
eigenvalues of Hamilton operator $H_k$.

For polynomial operators $N_{k}(\hat L_H,\hat R_{H})$ we have
\[ N_{k}(\hat L_H,\hat R_{H})  \rho=
\sum^{N}_{n=0} \sum^{n}_{m=0} a^{(k)}_{n,m} H^{m} \rho H^{n-m} . \]
In the general case, $m$ and $n$ are multi-indeces. The function
$N_{k}(E,E)$ is a polynomial
\[ N_{k}(E,E)=\sum^{N}_{n=0} \alpha^{(k)}_{n} E^{n} , \]
where
\[ \alpha^{(k)}_{n}=\sum^{n}_{m=0} a^{(k)}_{n,m} . \]

We can define the variable $x=E-a$, such that function
$N_k(E,E)=N_k(x+a,x+a)$ has no the term $x^{n-1}$.
\[ N_k(x+a,x+a)=\sum^{N}_{n=0} \alpha^{(k)}_{n} (x+a^{(k)})^{n}= \]
\[=\sum^{N}_{n=0} \sum^{n}_{m=0} \alpha^{(k)}_{n}
\frac{n!}{m!(n-m)!}x^{m}(a^{(k)})^{n-m} . \]
If the coefficient of the term $x^{n-1}$ is equal to zero
\[ \alpha^{(k)}_{n}\frac{n!}{(n-1)!}a^{(k)} +\alpha^{(k)}_{n-1}=
\alpha^{(k)}_{n} n a^{(k)} +\alpha^{(k)}_{n-1}=0  , \]
then we have
\[ a^{(k)}=-\frac{\alpha^{(k)}_{n-1}}{n \alpha^{(k)}_{n}}  . \]

If we change parameters $\alpha^{(k)}_n$, then can arise
stationary states of dissipative quantum systems.
For example, the bifurcation with birth of
linear oscillator stationary state is a quantum analog of
dynamical Hopf bifurcation \cite{TL,MMC}.

If the function $N(E,E)$ is equal to
\[ N(E,E)=\pm \alpha_{n}  E^n+ \sum^{n-1}_{j=1}
\alpha_{j} E^{j} \ \ \ n \ge 2 , \]
then potential $V(x)$ is 
\[ V(x)=\pm x^{n+1}+ \sum^{n-1}_{j=1}a_{j}x^{j} \quad n \ge 2 , \]
and we have catastrophe of type $A_{\pm n}$.

If we have $s$ variables $E_k$, where $k=1,2,...,s$, then quantum
analogous of elementary catastrophes $A_{\pm n}$, $D_{\pm n}$,
$E_{\pm 6}$, $E_7$ and $E_8$ can be realized. Let us write the
full list of typical set of potentials $V(x)$, which leads to
elementary catastrophes (zero-modal) defined by $V(x)=V_0(x)+Q(x)$, where
\[ A_{\pm n}: \ V_0(x)=\pm x^{n+1}_1+\sum^{n-1}_{j=1}a_{j}x^{j}_1 \ \
n \ge 2 , \qquad \qquad \quad \]
\[ D_{\pm n}: \ V_0(x)=x^{2}_{1}x_{2} \pm x^{n-1}_2+
\sum^{n-3}_{j=1}a_{j}x^{j}_{2}+\sum^{n-1}_{j=n-2}x^{j-(n-3)}_{1} , \]
\[ E_{\pm 6}: \ V_0(x)=(x^{3}_{1} \pm x^{4}_{2})+
\sum^{2}_{j=1}a_{j}x^{j}_{2}+\sum^{5}_{j=3}a_{j}x_{1}x^{j-3}_2 , \]
\[ E_{7}: \ V_0(x)=x^{3}_{1}+x_{1}x^{3}_{2}+
\sum^{4}_{j=1}a_{j}x^{j}_{2}+\sum^{6}_{j=5}a_{j}x_{1}x^{j-5}_2 , \]
\[ E_{8}: \ V_0(x)=x^{3}_{1}+x^{5}_{2}+
\sum^{3}_{j=1}a_{j}x^{j}_{2}+\sum^{7}_{j=4}a_{j}x_{1}x^{j-4}_2 . \]
Here $Q(x)$ is nondegenerate quadratic form with variables
$x_{2}$, $x_{3}$, ... , $x_{s}$ for $A_{\pm n}$ and parameters
$x_{3}$, ... , $x_{s}$ for other cases.

\section{Fold catastrophe}

In this section, we suggest an example of catastrophe $A_{2}$
called fold.

Let us consider Liouville-von Neumann equation for nonlinear
quantum oscillator with friction
\[ \frac{d}{dt}\rho_{t}=-\frac{i}{\hbar}[ H, \rho_{t}]+
\alpha_0 \frac{i}{\hbar}  [ q, p \circ  \rho_{t}]+ \]
\be \label{h3}
+\frac{i}{\hbar}\alpha_1[q, p \circ( H \circ  \rho_{t})]+
\frac{i}{\hbar}
\alpha_2 [q, p \circ( H \circ ( H \circ \rho_{t}))] , \ee
where $H$ is Hamilton operator defined by (\ref{harm}).

In this case, we have
\[ \hat F=\frac{i}{\hbar}(\hat L_q-\hat R_q)(\hat L_p+\hat R_p)  , \]
\[ N(\hat L_{H},\hat R_{H})= \alpha_0 \hat L_I+\frac{\alpha_1}{2}
(\hat L_{H}+\hat R_{H})+ \frac{\alpha_2}{4}(\hat L_{H}+\hat R_{H})^2 , \]
\[ N(E,E)=<\Psi|N( H, H)|\Psi>=\alpha_0+\alpha_1 E+\alpha_2 E^2 . \]
Stationary state $\rho_{\Psi}=|\Psi><\Psi|$ of harmonic oscillator
is stationary state of dissipative quantum system (\ref{h3}), if
\[ \alpha_0+\alpha_1  E+\alpha_2  E^2=0 . \]
If we define the variable $x$ and parameter $\lambda$ by
\[ x=E-a, \ \ a=-\frac{\alpha_{1}}{2\alpha_{2}} , \ \ \lambda=
\frac{4\alpha_{0}\alpha_{2}-\alpha^{2}_{1}}{4\alpha^{2}_{2}} , \]
then we have stationary condition $N(E,E)=0$ in the form
\[ x^{2}-\lambda=0 . \]
If $\lambda \le 0$, then  we have no stationary states.
If $\lambda >0$, then we have stationary states for discrete set
of parameter values $\lambda$.
If the parameters $a$ and $\lambda$ are equal to
\[ a=\frac{\hbar \omega}{2}(n_1+n_2+1) , \quad 
\lambda=\hbar^{2} \omega^{2} \frac{(n_1-n_2)^{2}}{4} , \]
where $n_{1}$ and $n_{2}$ are nonnegative integer numbers, then
dissipative quantum system has two stationary state (\ref{Psi}) of
linear harmonic oscillator. 
The energy of these states is equal to
\[ E_{n_1}=\hbar \omega(n_{1}+\frac{1}{2}) ,
\quad E_{n_2}=\hbar \omega(n_{2}+\frac{1}{2}) . \]

\section{Conclusion}

Dissipative quantum systems can have stationary states. Stationary
states of non-Hamiltonian and dissipative quantum systems can
coincide with stationary states of Hamiltonian systems. 
As an example we suggest quantum dissipative systems with
pure stationary states of linear harmonic oscillator.
Using (\ref{h1}), it is easy to get
dissipative quantum systems with stationary states of
hydrogen atom. For a special case of dissipative systems we can
use usual bifurcation  and catastrophe theory. It is easy to
derive quantum analogous of classical dynamical bifurcations.

Dissipative quantum systems with two stationary states
can be considered as qubit. It allows to consider
quantum computer with dissipation as
non-dissipative quantum computer.
In the general case, we can consider dissipative n-qubit quantum systems
as quantum computer with mixed states and quantum operations,
not necessarily unitary, as gates \cite{AKN,Tarpr}.
A mixed state (operator of density matrix) of n two-level quantum
systems is an element of $4^{n}$-dimensional operator Hilbert space.
It allows to use quantum computer 
model with 4-valued logic \cite{Tarpr}.
The gates of this model are general quantum operations which
act on the mixed state.

\section*{Acknowledgement}

This work was partially supported by the RFBR grant No. 02-02-16444.



\end{document}